\documentclass[aps, prd, showpacs, twocolumn, %showkeys, preprintnumbers,
amssymb,superscriptaddress]{revtex4}

\usepackage{graphicx, bm, amsmath, amsfonts}

\begin{document}

%\preprint{}

\title{
Composite-kink solutions of coupled 
nonlinear wave equations
}

\author{Hosho Katsura}
\affiliation{
Department of Physics, Gakushuin University, 
Mejiro, Toshima-ku, Tokyo 171-8588, Japan\\
}

\date{\today}
\begin{abstract}
We obtain exact traveling-wave solutions of the coupled nonlinear partial differential equations that describe the dynamics of two classical scalar fields in 1+1 dimensions. The solutions are kinks interpolating between neighboring vacua. We compute the classical kink mass and show that it saturates a Bogomol'nyi-type bound. We also present exact traveling-wave solutions of a more general class of models. Examples include coupled $\phi^4$ and sine-Gordon models. 
\end{abstract}

% 11.27.+d	Extended classical solutions; cosmic strings, domain walls, texture
% 03.50.Kk	Other special classical field theories
% 11.10.Lm	Nonlinear or nonlocal theories and models

\maketitle
\section{Introduction}

Nonlinear partial differential equations, particularly those exhibiting solitary-wave solutions, have been of great interest for many decades. Their applications range from classical and quantum field theories to condensed-matter problems~\cite{Bishop, 1d_physics}. 
Typical examples are the equations of motion for a single scalar field in 1+1 dimensions. Tremendous efforts have been made to find solitary-wave solutions in these systems. For completely integrable systems, like the sine-Gordon model, one can apply a variety of sophisticated techniques such as the B\"acklund transformation~\cite{Lamb}. A more down-to-earth approach, called the Bogomol'nyi method~\cite{Bogo, kinks_dw}, can be applied to even non-integrable models. This method reduces the original second-order differential equation to a first-order one, solutions of which are known as kinks (or domain walls)~\cite{kinks_dw}. 
%and minimize the energy with fixed boundary conditions. 
Kinks are perhaps the simplest kinds of topological solitons~\cite{Rajaraman, Manton}. They behave as particle-like objects, with finite mass and a smooth structure. 
%Also, they can give important insights into nonperturbative aspects of quantum field theories~\cite{Frishman, Samaj}. Examples in which such kinks appear are the sine-Gordon and $\phi^4$ models in 1+1 dimension. 

At the next level of complexity, one can consider systems of two coupled scalar fields in 1+1 dimensions. It is also natural to ask whether topological solitons are special solutions of the equations of motion in such cases. Rajaraman proposed a method for finding exact solutions of the problems~\cite{Rajaraman_PRL}. His procedure is model-independent and can be, in principle, used for any model. In practice, however, the application of his method is limited since there is a need to choose ``trial orbits" on which the two fields satisfy some constraint. 
%\textbf{[H.K.: Should I touch on old Russian papers and Saxena's papers?]} 
Bazeia {\it et al.} proposed another general method for constructing models exhibiting soliton solutions~\cite{Bazeia1, Bazeia2, Bazeia3, Bazeia4}. Their approach is based on the idea of the Bogomol'nyi method. They also provided specific examples including coupled $\phi^4$ models. 
%~\cite{SY_You_99}

In this paper, we present a new class of systems with exact kink solutions 
%study systems of two coupled scalar fields 
that do not belong to the class of models in Refs. \cite{Bazeia1, Bazeia2, Bazeia3, Bazeia4}. Our approach is %based on a combination 
a hybrid of Rajaraman's and Bogomol'nyi's methods. Namely, the constraint on the two fields is embedded in a potential term, which does not alter the Bogomol'nyi bound. The simplest example 
%is a model of coupled $\phi^4$- and sine-Gordon models. 
presented in the next section can be thought of as an effective field theory describing the coupling of magnetic and ferroelectric domain walls in multiferroic GdFeO$_3$~\cite{Furukawa-Katsura, multiferroics}, where the composite of these domain walls has been observed in experiments~\cite{Tokunaga_09}. 
We show that when the parameter of the model is fine-tuned to a particular value, the composite of kinks in the individual scalar field theories, which we call the composite kink, is an exact solution of the equations of motion. We then 
generalize the idea of Bazeia {\it et al.}, and show that the mass of the composite kink saturates a Bogomol'nyi-type bound. At the end of the paper, we provide a general recipe for constructing models in which the composite kinks are exact solutions of the equations of motion. 
%Examples include coupled $\phi^4$ and sine-Gordon models. 

\section{Model}
\label{sec:model}
\begin{figure}
\centering
\includegraphics[width=0.55\columnwidth]{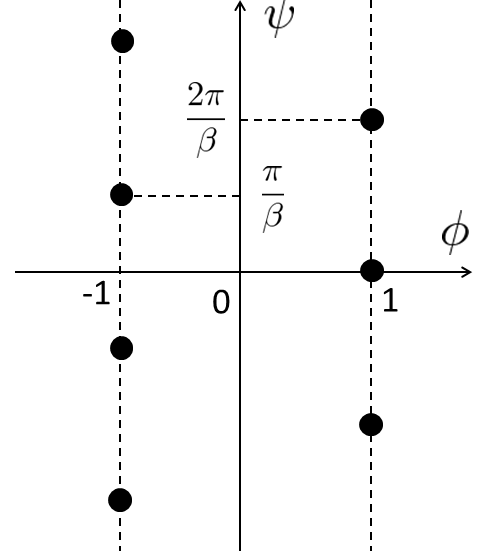}
\caption{Locations of the minima of the potential $U(\phi, \psi)$, which form a regular lattice.}
\label{fig:vacua_configuration}
\end{figure}

Consider two real scalar fields $\phi=\phi(x,t)$ and $\psi = \psi (x,t)$. In the following, we use the standard notation $x^\alpha = (t,x)$, $x_\alpha=(t,-x)$ and natural units $\hbar=c=1$. In our model, the Lagrangian density 
%describing the dynamics of $\phi$ and $\psi$ 
is given by 
\begin{eqnarray}
&& {\cal L} = \frac{1}{2}\partial_\alpha \phi\, \partial^\alpha \phi 
+ \frac{1}{2}\partial_\alpha \psi\, \partial^\alpha \psi-U(\phi,\psi),
\label{eq:pre-lagran}\\
&& U(\phi,\psi) = \frac{\lambda}{4} \phi^4 -\mu \phi \cos (\beta \psi) + U_0, 
\end{eqnarray}
where $\lambda>0$, $\mu$, $\beta$, and $U_0$ are real numbers. Note that $\mu$ can be either positive or negative. However, the Lagrangian density with negative $\mu$ can be obtained from that with positive $\mu$ by changing $\phi \to -\phi$. Thus, we may assume that $\mu>0$ without loss of generality. The constant $U_0$ is determined so that the classical vacuum energy is zero. The Lagrangian density ${\cal L}$ is invariant under the combination of reflection $\phi \to -\phi$ and translation $\psi \to \psi + \pi n/\beta$ ($n=$ odd integers). One can simplify ${\cal L}$ by introducing scaled variables, 
\begin{equation}
({\bar x}, {\bar t}) = (x/A, t/A),~~
({\bar \phi}, {\bar \psi}) = (\phi/B, \psi/B), 
\end{equation}
with $A=\lambda^{-1/6} \mu^{-1/3}$ and $B=\lambda^{-1/3} \mu^{1/3}$. In terms of these variables, the Lagrangian density is written as 
\begin{eqnarray}
{\cal L} \!=\! \lambda^{\!-\!1/3}\!\mu^{\!4/3\!} \! \left[
\!\frac{1}{2}\partial_\alpha {\bar \phi}\, \partial^\alpha {\bar \phi} 
\!+\!\frac{1}{2}\partial_\alpha {\bar \psi}\, \partial^\alpha {\bar \psi}
\!-\!\frac{{\bar \phi}^4}{4}\!+\!{\bar \phi} \cos ({\bar \beta} {\bar \psi}) 
\!-\! {\bar U}_0 \!\right]\!,
\nonumber
\end{eqnarray}
where ${\bar \beta} = B \beta$ and ${\bar U}_0= \lambda^{1/3}\mu^{-4/3} U_0$. Thus, without loss of generality, one can set $\lambda=\mu=1$. In the following, we drop the bars over the scaled variables and simply write 
\begin{eqnarray}
&& {\cal L} = \frac{1}{2}\partial_\alpha \phi\, \partial^\alpha \phi 
+ \frac{1}{2}\partial_\alpha \psi\, \partial^\alpha \psi - U(\phi,\psi),
\label{eq:lagran}\\
&& U(\phi,\psi) = \frac{1}{4} \phi^4 -\phi \cos (\beta \psi) + U_0.
\label{eq:pot0}
\end{eqnarray}
The potential term $U(\phi,\psi)$ has an infinite series of degenerate minima placed at 
$(\phi, \psi) = (+1, 2\pi n/\beta)$ and $(\phi, \psi) = (-1, 2\pi (n+1/2)/\beta)$ ($n \in \mathbb{Z}$), 
corresponding to the classical vacua of the field theory described by ${\cal L}$ (see Fig. \ref{fig:vacua_configuration}). The requirement $U(\phi, \psi)=0$ at the vacuum configurations yields $U_0 = 3/4$.

%%%%%%%%%% Figure :  vacua cofiguration %%%%%%%%%%
%%%%%%%%%

\section{Exact solutions}
The equations of motion for $(\phi, \psi)$ can be derived from the Euler-Lagrange equations
\begin{equation}
\partial_\mu \left( \frac{\partial {\cal L}}{\partial (\partial_\mu \phi)} \right) - \frac{\partial {\cal L}}{\partial \phi}=0,~~
\partial_\mu \left( \frac{\partial {\cal L}}{\partial (\partial_\mu \psi)} \right) - \frac{\partial {\cal L}}{\partial \psi}=0
\end{equation}
with Eqs. (\ref{eq:lagran}, \ref{eq:pot0}). They read 
\begin{eqnarray}
%&& \partial^2_t \phi -\partial^2_x \phi 
&& \partial_\alpha \partial^\alpha \phi +\phi^3 - \cos (\beta \psi) = 0,
\label{eq:phi4_1}\\
%&& \partial^2_t \psi -\partial^2_x \psi 
&& \partial_\alpha \partial^\alpha \psi + \beta \phi \sin (\beta \psi) = 0.
\label{eq:SG1}
\end{eqnarray}
It seems quite unlikely that the above coupled nonlinear equations allow us 
to find exact solutions. However, we will show that kinks that interpolate 
between neighboring vacua are exact traveling-wave solutions of 
Eqs. (\ref{eq:phi4_1}, \ref{eq:SG1}) when the parameter is set to $\beta=1/\sqrt{2}$. 

%%%%%%%%%% Figure :  kinks_ab %%%%%%%%%%
\begin{figure}
\centering
\includegraphics[width=0.75\columnwidth]{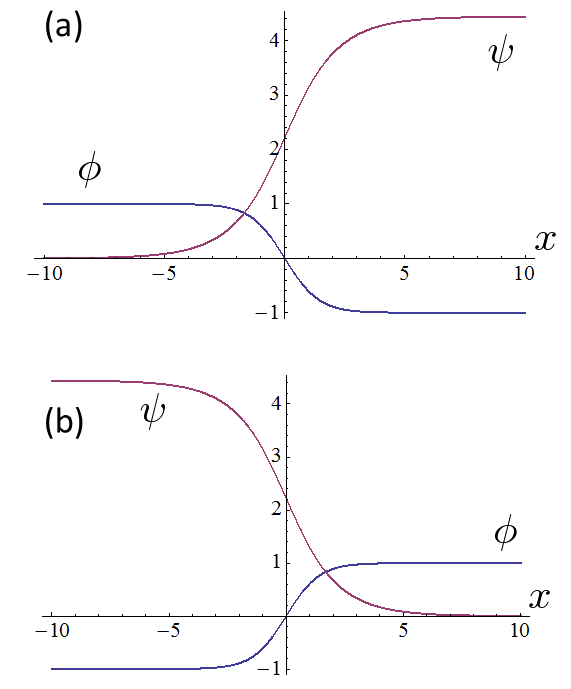}
\caption{Kinks interpolating between (a) $(\phi, \psi)=(1, 0)$ and $(-1, \pi/\beta)$, and (b) $(-1, \pi/\beta)$ and $(1,0)$. 
}
\label{fig:kinks_ab}
\end{figure}
%%%%%%%%%

Let us first find static solutions. Since the model is Lorentz invariant, 
if one finds static solutions, one can Lorentz boost them and get traveling-wave solutions. 
The equations to be solved are
\begin{eqnarray}
\frac{d^2 \phi}{dx^2} &=& \frac{\partial U}{\partial \phi} = \phi^3 - \cos (\beta \psi),
\\
\frac{d^2\psi}{dx^2} &=& \frac{\partial U}{\partial \psi} = \beta \phi \sin (\beta \psi).
\label{eq:SG2}
\end{eqnarray}
To find special solutions of the above equations, we use the following ansatz:
\begin{equation}
\phi = C \cos (\beta \psi),
\label{eq:ansatz}
\end{equation}
where $C$ is a constant. With this ansatz the equations become 
\begin{eqnarray}
&& \frac{d^2\phi}{dx^2} = \phi^3 - \frac{\phi}{C},
\label{eq:phi4_3}\\
&& \frac{d^2 \psi}{dx^2} = \frac{\beta}{2} C \sin (2 \beta \psi),
\label{eq:SG3}
\end{eqnarray}
where $\phi$ and $\psi$ are formally decoupled. Now recall that static kink (anti-kink) solutions of the $\phi^4$ and the sine-Gordon models, respectively, satisfy Eqs. (\ref{eq:phi4_3}) and (\ref{eq:SG3}). The solutions are given by
\begin{eqnarray}
 \phi^\pm (x) \!=\! \pm \frac{1}{\sqrt{C}} \tanh \left[ \frac{1}{\sqrt{2C}} (x-X_1) \right],
\label{eq:kink1}\\
\!\!\!\! \psi^\pm (x) \!=\! \pm \frac{1}{\beta} \arccos \left\{
\pm \tanh \left[ \beta \sqrt{C}  (x-X_2) \right]
\right\} \!+\! \frac{2\pi}{\beta} n,
%\cos (\beta \psi (x)) = \pm \tanh \left[ \beta \sqrt{C}  (x-X_2) \right],
\label{eq:kink2}
\end{eqnarray}
where $X_1$ and $X_2$ denote the positions of the kink centers, and $n \in \mathbb{Z}$. Here the principal value of $\arccos$ is used. 
The parameters $\beta$, $C$, $X_1$, and $X_2$ are determined so that $\phi^\pm(x)$ and $\psi^\pm(x)$ satisfy the condition (\ref{eq:ansatz}). 
Substituting Eqs. (\ref{eq:kink1}, \ref{eq:kink2}) into Eq. (\ref{eq:ansatz}), we find $\beta=1/\sqrt{2}$ and $C=1$. $X_1$, $X_2$ can be arbitrary as long as $X_1=X_2$. Examples of the solutions are shown in Fig. \ref{fig:kinks_ab}. 
From the static solutions Eqs. (\ref{eq:kink1}, \ref{eq:kink2}), traveling-wave solutions can be constructed as $\phi^\pm (x,t) = \phi^\pm (\gamma (x-vt))$, $\psi^\pm (x,t) = \psi^\pm (\gamma (x-vt))$, where $\gamma=1/\sqrt{1-v^2}$. 
Thus we see that the composite of kinks in the $\phi^4$ and sine-Gordon models, which we call composite kink, is an exact solution of the coupled model. 
The field configurations at $|x| \to \infty$ are
\begin{eqnarray}
&& (\phi^+ (\infty), \psi^+ (\infty) ) = (+1, 2\pi n/\beta), \\
&& (\phi^+ (-\infty), \psi^+ (-\infty) ) = (-1, 2\pi(n+1/2)/\beta), \\
&& (\phi^- (\infty), \psi^- (\infty) ) = (-1, 2\pi(n+1/2)/\beta), \\
&& (\phi^- (-\infty), \psi^- (-\infty) ) = (+1, 2\pi n/\beta),
\end{eqnarray}
with $\beta=1/\sqrt{2}$. It is thus clear that the composite kinks interpolate between neighboring vacua (see Fig. \ref{fig:kinks_ab}).

\section{Kink mass}
In this section, we focus on the exactly solvable case where the parameter is set equal to $\beta=1/\sqrt{2}$, and compute the total energy of the composite kink, 
i.e., the kink mass. %static kink solutions, which is sometimes called kink mass. 
The exact static solutions of Eqs. (\ref{eq:phi4_1}, \ref{eq:SG1}) are 
\begin{eqnarray}
\phi^\pm_{0} (x) &\!=\!& \pm \tanh \left( \frac{x}{\sqrt{2}} \right),~
\label{eq:stat_sol1} \\
\psi^\pm_{0} (x) &\!=\!& 
\pm \sqrt{2} \arccos \left[
\pm \tanh \left( \frac{x}{\sqrt{2}} \right)
\right] \!+\! 2\sqrt{2} \pi n,
\label{eq:stat_sol2}
%\cos \left( \frac{\psi_{\rm kink} (x)}{\sqrt{2}} \right) = \pm \tanh \left( \frac{x}{\sqrt{2}} \right),
\label{eq:stat_sol2}
\end{eqnarray}
where $n \in \mathbb{Z}$ and we choose the center of the kinks at the origin ($x=0$). 
We will show that these solutions saturate a Bogomol'nyi-type bound. 

For a general solution of the system given by Eqs. (\ref{eq:lagran}) and (\ref{eq:pot0}), 
the total energy can be written as 
\begin{eqnarray}
E = \int^\infty_{-\infty} &\bigg[& \!
 \frac{1}{2}\left( \frac{\partial \phi}{\partial t} \right)^2 \!
+\frac{1}{2}\left( \frac{\partial \phi}{\partial x} \right)^2\!
\nonumber \\
&+&\! \frac{1}{2}\left( \frac{\partial \psi}{\partial t} \right)^2 \!
+\frac{1}{2}\left( \frac{\partial \psi}{\partial x} \right)^2
+ U(\phi, \psi)
\bigg],
\end{eqnarray} 
from which it is clear that the total energy is bounded from below by those of static solutions. Thus we have
%The total energy of the static solution, sometimes called kink mass, is 
\begin{equation}
E \ge \int^\infty_{-\infty} \left[
  \frac{1}{2}\left( \frac{d\phi}{dx} \right)^2 
+\frac{1}{2}\left( \frac{d\psi}{dx} \right)^2 + U(\phi, \psi)
\right], 
\label{eq:ineq1}
\end{equation}
where $\phi$ and $\psi$ denote static solutions. Then from Eq. (\ref{eq:pot0}), we notice that the potential at the solvable point ($\beta=1/\sqrt{2}$) can be written as
\begin{equation}
U(\phi, \psi) = \frac{1}{2} \left( \frac{\partial W}{\partial \phi} \right)^2 
+\frac{1}{2}\left( \frac{\partial W}{\partial \psi} \right)^2
+\frac{1}{2} V^2,
%+\frac{1}{2} \left( \phi- \cos \left( \frac{\psi}{\sqrt{2}} \right) \right)^2,
\label{eq:pot1}
\end{equation}
where
\begin{eqnarray}
W (\phi, \psi) &=& \frac{1}{\sqrt{2}} \left[ -\frac{\phi^3}{3} + \phi +2 \cos \left( \frac{\psi}{\sqrt{2}} \right) \right],~
\label{eq:super} \\
V (\phi, \psi) &=& \phi- \cos \left( \frac{\psi}{\sqrt{2}} \right).
\label{eq:Vpot}
\end{eqnarray}
%Then it is easy to see that the total energy can be rewritten as 
Substituting them into Eq. (\ref{eq:ineq1}), we find
\begin{eqnarray}
E \ge \!\!\!\!\! &&\frac{1}{2} \int^\infty_{-\infty} \left[\!
 \left(\! \frac{d\phi}{dx} \right)^2 \!\!+\! \left( \frac{d\psi}{dx} \!\right)^2 
\!\!+\! \left(\! \frac{\partial W}{\partial \phi} \!\right)^2 \!\!+\! \left(\! \frac{\partial W}{\partial \psi} \!\right)^2
\!+\!V^2
\right]
\nonumber \\
= \!\!\!\!\! &&\frac{1}{2} \int^\infty_{-\infty} \left[\!
  \left( \frac{d\phi}{dx} \mp \frac{\partial W}{\partial \phi} \right)^2 
\!\!+\! \left( \frac{d\psi}{dx} \mp \frac{\partial W}{\partial \psi} \right)^2 
\!\!+\! V^2 
\right]
\nonumber \\
&&\pm \int^\infty_{-\infty} \left(
 \frac{\partial W}{\partial \phi}\, \frac{d\phi}{dx}
+\frac{\partial W}{\partial \psi}\, \frac{d\psi}{dx}
\right).
\label{eq:BPS1}
\end{eqnarray}
Now recall that
\begin{equation}
\frac{dW}{dx} = \frac{\partial W}{\partial \phi}\, \frac{d\phi}{dx}
+\frac{\partial W}{\partial \psi}\, \frac{d\psi}{dx},
\end{equation}
and the integrand in the second line of Eq. (\ref{eq:BPS1}) is a sum of squares, 
we see that the total energy is bounded from below by
\begin{equation}
M = |W(\phi(\infty), \psi(\infty)) - W(\phi(-\infty),\psi(-\infty))|. 
%= \frac{8\sqrt{2}}{3}. 
\end{equation}
Here the energy bound is obtained solely from topological data (field configurations at $|x| \to \infty$). Energy bounds of this type are called Bogomol'nyi bounds. 
The minimal energy configuration $E=M$ is achieved when static solutions $\phi(x)$ and $\psi(x)$ satisfy the following first-order equations and constraint:
\begin{equation}
\frac{d\phi}{dx} = \pm \frac{\partial W}{\partial \phi},~
\frac{d\psi}{dx} = \pm \frac{\partial W}{\partial \psi},~
V(\phi, \psi) = 0.
\label{eq:first}
\end{equation}
One can verify that the solutions $\phi^\pm_{0}(x)$ and $\psi^\pm_{0} (x)$ in Eqs. (\ref{eq:stat_sol1}, \ref{eq:stat_sol2}) satisfy the above conditions. The condition $V (\phi^\pm_{0}, \psi^\pm_{0})=0$ follows from the fact that the special solutions satisfy the ansatz Eq. (\ref{eq:ansatz}). It should be noted that $(\phi^\pm_{0}(x), \psi^\pm_{0} (x))$ is the unique solution (up to translation) that minimizes the total energy with kink boundary conditions. 
Now the total energy %of the static kink solutions 
is computed as
\begin{eqnarray}
M &\!=\!& \int^\infty_{-\infty}\!\! \left[
  \frac{1}{2}\left(\! \frac{d\phi^\pm_{0}}{dx} \!\right)^2 \!+
\frac{1}{2}\left(\! \frac{d\psi^\pm_{0}}{dx} \!\right)^2 \!+ U(\phi^\pm_{0}, \psi^\pm_{0})
\right]
\nonumber \\
&\!=\!&|W(\phi^\pm_{0} (\infty), \psi^\pm_{0} (\infty)) - W(\phi^\pm_{0} (-\infty),\psi^\pm_{0} (-\infty))|
\nonumber \\
&\!=\!& \frac{8 \sqrt{2}}{3}. 
\end{eqnarray}
Note that the masses of the $\phi^4$ and the sine-Gordon kinks are, respectively, $M_{\phi 4}=2\sqrt{2}/3$ and $M_{\rm SG}=2\sqrt{2}$, which add up to $M$. 
%\textbf{[H.K.: Is it a sum of individual kink masses?]}

\section{Generalized model}
The procedure presented in the previous section suggests that there is a large class of models in which the composite kinks are special solutions of the equations of motion. For example, one can consider the following one-parameter deformation of the potential:
\begin{equation}
U (\phi, \psi; d) = \frac{1}{2} \left( \frac{\partial W}{\partial \phi} \right)^2 
+\frac{1}{2}\left( \frac{\partial W}{\partial \psi} \right)^2
+\frac{d^2}{2} V^2,
\end{equation}
where $d$ is real, and $W(\phi, \psi)$ and $V(\phi,\psi)$ are defined in Eqs. (\ref{eq:super}, \ref{eq:Vpot}). 
The corresponding Lagrangian density reads
\begin{eqnarray}
{\cal L}_d &=& {\cal L}_{\phi 4} + {\cal L}_{\rm SG} + {\cal L}_{\rm int}, \\
{\cal L}_{\phi 4} &=& \frac{1}{2} \partial_\alpha \phi\, \partial^\alpha \phi 
-\frac{\phi^4}{4} +\frac{1-d^2}{2} \phi^2, \\
{\cal L}_{\rm SG} &=& \frac{1}{2} \partial_\alpha \psi\, \partial^\alpha \psi 
+\frac{1-d^2}{4} \cos (2 \beta \psi), \\
{\cal L}_{\rm int} &=& d^2 \phi \cos (\beta \psi) - \frac{2+d^2}{4}, 
\end{eqnarray}
with $\beta =1/\sqrt{2}$. %One can interpret the model 
The model can be interpreted as coupled $\phi^4$ and sine-Gordon models. The Lagrangian Eq. (\ref{eq:lagran}) with the potential Eq. (\ref{eq:pot0}) is the limiting case of ${\cal L}_d$ ($d=1$).
A similar but not identical model was presented in Rajaraman's work~\cite{Rajaraman_PRL}. However, we emphasize that our model is more realistic than that model, which is somewhat unphysical in the sense that the coupling of the quartic term is negative and hence unbounded from below. In addition, our model can be thought of as an effective field theory describing composite kinks in multiferroic GdFeO$_3$~\cite{Furukawa-Katsura, Tokunaga_09}. 

The traveling-wave solutions obtained by Lorentz boosting $\phi^\pm_0 (x)$ and $\psi^\pm_0 (x)$ in Eqs. (\ref{eq:stat_sol1}, \ref{eq:stat_sol2}) are still solutions of the equations of motion derived from the deformed Lagrangian density ${\cal L}_d$. 
This can be seen by noting that the first-order differential equations and the constraint
\begin{equation}
\frac{d \phi}{dx} = \pm \frac{\partial W}{\partial \phi},~
\frac{d \psi}{dx} = \pm \frac{\partial W}{\partial \psi},~
V (\phi, \psi) = 0,
\end{equation}
solve the equations for static solutions 
\begin{eqnarray}
\frac{d^2 \phi}{dx^2} &=& \frac{\partial^2 W}{\partial \phi^2}\, \frac{\partial W}{\partial \phi}
+ d^2 V \frac{\partial V}{\partial \phi},
\\
\frac{d^2 \psi}{dx^2} &=& \frac{\partial^2 W}{\partial \psi^2}\, \frac{\partial W}{\partial \psi}
+ d^2 V \frac{\partial V}{\partial \psi}, 
\end{eqnarray}
where we have used the fact that $\partial^2 W/\partial \phi \partial \psi=0$. Repeating the argument in the previous section, one can show that the kink mass 
$M$ does not depend on the deformation parameter $d$, i.e., $M=8\sqrt{2}/3$. 

Since the detailed profile of $V(\phi, \psi)$ is not essential in the proof, 
we come up with a more general model in which the potential has $n$ real deformation parameters $d_1, ..., d_n$:
\begin{eqnarray}
U (\phi, \psi; d_1,...,d_n) &\!=\!&\frac{1}{2} \left( \frac{\partial W}{\partial \phi} \right)^2 
+\frac{1}{2}\left( \frac{\partial W}{\partial \psi} \right)^2
\nonumber \\
&\!+\!& \frac{1}{2} \sum^n_{k=1} d^2_k \left[ \phi - \cos \left( \frac{\psi}{\sqrt 2} \right) \right]^{2k}\!\!\!\!. 
\end{eqnarray}
By construction, the potential is non-negative. 
Along the same line of reasoning, one can show that the traveling-wave solutions 
obtained by Lorentz boosting $\phi^\pm_0 (x)$ and $\psi^\pm_0 (x)$ in Eqs. (\ref{eq:stat_sol1}, \ref{eq:stat_sol2}) are 
exact solutions of the equations of motion. The models with $U(\phi, \psi; d_1, ..., d_n)$ can be thought of as coupled $\phi^{2n}$ and multi-frequency sine-Gordon models. 

\section{Concluding remarks}
In this paper, we have investigated a class of coupled nonlinear partial differential equations that describe the dynamics of two real scalar fields in 1+1 dimensions. 
The simplest model presented in Sec. \ref{sec:model} contains only a single parameter $\beta$. We have shown that when this parameter is fine-tuned to $\beta=1/\sqrt{2}$, the composite kink (a bound state of kinks) is an exact solution of the equations of motion. We also found that a variant of the Bogomol'nyi method greatly simplifies the computation of the mass of the composite kinks. This hints at the existence of a large class of models in which the composite kinks are exact solutions. We showed, by example, how one can construct such models. Note that generalizations of the method to $N$ coupled fields is quite straightforward. One can even consider a quasi-one-dimensional case, where the coupled one-dimensional systems form a lattice in the perpendicular plane. 

It is natural to ask whether the composite-kink solutions we found are stable against perturbations. The standard approach is to study the linear stability of the solutions by solving one-dimensinal Schr\"odinger equations whose (matrix) potential is determined by the spatial profile of the composite kinks~\cite{Bazeia_stability}. 
If the composite-kink solutions are stable, it would be interesting to consider a quantization of them and compute the quantum corrections to the kink masses~\cite{kinks_dw, Lohe, Izquierdo}. These issues will be discussed elsewhere~\cite{Sakamoto}. 

%%%%%%%%%%%%%%%%%%%%%%%%%%%%%%%%%%%%%%%
%%%%%%       ACKNOWLEDGMENT       %%%%%%
%%%%%%%%%%%%%%%%%%%%%%%%%%%%%%%%%%%%%%%
\section*{Acknowledgment}
The author thanks Ryohei Sakamoto for valuable discussions and for pointing out some errors in the earlier version of the paper. The author also thanks Nobuo Furukawa and Tohru Koma for valuable discussions. This work was supported by Grant-in-Aid 
%for JSPS Fellows (23-7601) and 
for Young Scientists (B) (23740298). 

%\appendix
%\section{Zero-energy ground state}


\begin{thebibliography}{99}
\bibitem{Bishop}
{\it Solitons and Condensed Matter Physics}, 
edited by A. R. Bishop and T. Schneider (Springer-Verlag, Berlin, 1978).

\bibitem{1d_physics}
{\it Physics in One Dimension}, 
edited by J. Bernasconi and T. Schneider (Springer, Berlin, 1981).

\bibitem{Lamb}
G. L. Lamb, {\it Elements of Soliton Theory} (John Wiley and Sons, Inc., New York, 1980).

\bibitem{Bogo}
E. B. Bogomol'nyi, Sov. J. Nucl. Phys. \textbf{24}, 449 (1976).

\bibitem{kinks_dw}
T Vachaspati, {\it  Kinks and Domain Walls} 
(Cambridge University Press, Cambridge, England, 2006).

\bibitem{Rajaraman}
R. Rajaraman, {\it Solitons and Instantons: An Introduction to
Solitons and Instantons in Quantum Field Theory} 
(North-Holland, Amsterdam, 1982).

\bibitem{Manton}
N. Manton and P. Sutcliffe, {\it Topological Solitons} 
(Cambridge University Press, Cambridge, England, 2004).

%\bibitem{Frishman}
%\bibitem{Samaj}

\bibitem{Rajaraman_PRL}
R. Rajaraman, Phys. Rev. Lett. \textbf{42}, 200 (1979).

%\bibitem{SY_You_99}
%S.-Y. Lou, J. Phys. A {\textbf 32}, 4521 (1999).
%S.-Y. Lou, J. Phys. A {\textbf 32}, 4521 (1999).

\bibitem{Bazeia1}
D. Bazeia, M. J. dos Santos, R. F. Ribeiro, Phys. Lett. A \textbf{208}, 84 (1995), {\tt arXiv:hep-th/0311265}.
\bibitem{Bazeia2}
D. Bazeia and M. M. Santos, Phys. Lett. \textbf{217A}, 28 (1996).
\bibitem{Bazeia3}
D. Bazeia, R. F. Ribeiro and M. M. Santos, Phys. Rev. E \textbf{54}, 2943 (1996).
\bibitem{Bazeia4}
D. Bazeia, R. F. Ribeiro, and M. M. Santos, Phys. Rev. D \textbf{54}, 1852 (1996).

\bibitem{Furukawa-Katsura}
N. Furukawa and H. Katsura, "Composite Domain Walls in Multiferroic Orthoferrites RFeO3." Bulletin of the American Physical Society \textbf{54} (2009). 

\bibitem{multiferroics}
For a review of multiferroic materials, see K. F. Wang, J. M. Liu, and Z. F. Ren, Adv. Phys. \textbf{58}, 321 (2009), {\tt arXiv:cond-mat/0908.0662}.

\bibitem{Tokunaga_09}
Y. Tokunaga, N. Furukawa, H. Sakai, Y. Taguchi, T. Arima and Y. Tokura, 
Nature Mater. 8, 558 (2009).

\bibitem{Bazeia_stability}
D. Bazeia, J. R. S. Nascimento, R. F. Ribeiro, and D. Toledo, 
J. Phys. A \textbf{30}, 8157 (1997), {\tt arXiv:hep-th/9705224}.

\bibitem{Lohe}
M. A. Lohe, Phys. Rev. D \textbf{20}, 3120 (1979).

\bibitem{Izquierdo}
A. A. Izquierdo, W. G. Fuertes, M. A. Leon, M. Mayado, J. M. Guilarte, and J. M. Castaneda, {\it Lectures on the mass of topological solitons} (2006), {\tt arXiv:hep-th/0611180}.

\bibitem{Sakamoto}
H. Katsura and R. Sakamoto, unpublished.

\end{thebibliography}
\end{document}